\documentstyle[12pt,fullpage,psfig]{article}
\def\beq{\begin{equation}}
\def\eeq{\end{equation}}

\begin{document}
\begin{center}
{\Large{\bf The Abelian Sandpile and Related Models}}\\[2cm]

{\large{\bf Deepak Dhar}}\\
Department of Theoretical Physics, \\ 
Tata Institute of Fundamental Research, \\ 
Homi Bhabha Road, Mumbai 400~005, INDIA\\ [2cm]
\underbar{Abstract}
\end{center}

The Abelian sandpile model is the simplest analytically tractable model of
self-organized criticality.  This paper presents a brief review of known
results about the model. The abelian group structure of the algebra of
operators allows an exact calculation of many of its properties. In
particular, when there is a preferred direction, one can calculate all the
critical exponents characterizing the distribution of avalanche-sizes in
all dimensions. For the undirected case, the model is related to $q
\rightarrow 0$ state Potts model. This enables exact calculation of some
exponents in two dimensions, and there are some conjectures about others.
We also discuss a generalization of the model to a network of
communicating reactive processors. This includes sandpile models with
stochastic toppling rules as a special case. We also consider a
non-abelian stochastic variant, which lies in a different universality
class, related to directed percolation.

\section{ Introduction}

It has been about 10 years since Bak, Tang and Wiesenfeld's landmark
papers on self-organized criticality appeared \cite{BTW1,BTW2}. In this
period, the concept of self-organized criticality has been invoked to
describe a large variety of different systems such as forest-fires,
earthquakes, punctuated equilibrium in biology, stock-market fluctuations
etc.. I shall not attempt to review the very large number of papers
inspired by Bak's justly influential ideas. A readable overview may be
found in his recent book \cite{bak}.  Instead I shall concentrate on one
specific model: the abelian sandpile model. Even here, rather than provide
a summary of all the papers on this problem, I will try to provide an
overview, from a perhaps somewhat biassed personal perspective. The review
is not self-contained, and for details of arguments the reader will have
to consult the original papers. Even so, I hope that it will be a useful
introduction, and guide to literature, for students and others trying to
learn about this subject for the first time. The paper also contains some
unpublished material (mainly in section 5). Another recent review, similar
in scope, is by Ivashkevich and Priezzhev \cite{ivashrev}.  Brief accounts
of my own work on this model have appeared in conference proceedings
earlier\cite{ddrev1,ddrev2}. 

The sandpile model was proposed as a paradigm of self-organized
criticality (SOC). It is certainly the simplest, and best understood, {\it
theorist's} model of SOC: it is a non-equilibrium system, driven at a slow
steady rate, with local threshold relaxation rules, which in the steady
state shows relaxation events in bursts of a wide range of sizes, and
long-range spatio-temporal correlations, obtained without fine-tuning of
any control parameters. 

The Abelian sandpile model (ASM) is the name given to a particular
subclass of the sandpile models that have a nice mathematical structure
(an abelian group). The group structure allows analytical calculation of
many of the properties of the model. The mathematical tractability of the
model is certainly its main attraction. In addition, the model turns out
to be related to several other models in statistical mechanics: the Potts
model, the voter model, directed percolation, Takayasu aggregation model
etc.. 

A different reason for the continued interest of physicists in the model
is its intractability: in spite of its apparent simplicity, the original
2-dimensional BTW model has defied an analytical calculation of all
critical exponents so far. The fact that the steady state for the ASM is
well characterized, and allows exact calculation of averages of many
physical quantities of interest ( and with only moderate effort), and that
several critical exponents are known exactly, suggests that the model is
`soluble', and other exponents can also be determined analytically (if
only we could figure out how!). 

\section{ General properties of the Abelian Sandpile Model}

The ASM is defined as follows \cite{dd90}: we consider a lattice of $N$
sites labelled by integers $i=1$ to $N$. At each site $i$, we define a
nonnegative integer height variable $z_i$ called the height of the
sandpile, and a threshold value $z_{ic}$. If $z_i < z_{ic}$, for all $i$,
the pile is said to be stable. The time-evolution of the sandpile is
defined by the following two rules:

1. Adding a particle:  Select one of the sites randomly, the probability
that site $i$ is picked being some given value $p_i$, and we add a grain
of sand there.  Clearly, $\sum_i p_i=1$. On addition of the grain at site
$i$, $z_i$ increases by 1. Height at other sites remains unchanged. 

2. Toppling: If for any site $z_i \geq z_{ic}$, then that site is said to
be unstable, it topples, and loses some sandgrains to other sites.  This
sandgrain transfer is defined in terms of an $N \times N$ integer matrix
$\Delta$. On toppling at site $i$ , the configuration is updated according
to the rule: 

\beq
 z_j \rightarrow z_j -\Delta_{ij}, for ~j =~1~ to~ N.  
\eeq

If the toppling results in some other site becoming unstable, it is also
toppled. The process continues until all sites become stable. 

The matrix $\Delta$ is assumed to have the following properties: 

 i) $\Delta_{ii} > 0,$ for all $i$.

 ii) $\Delta _{ij} \leq 0$, for all $i \neq j$.
 
 iii) $\sum_j \Delta_{ij} \geq 0$, for all $i$.

These conditions just ensure that on toppling at site $i$, $z_i$ must
decrease, and height at other sites $j$ can only increase, and there is no
creation of sand in the toppling process. Some sand may get lost from the
system if the toppling occurs at a boundary site. In fact, no stationary
state of the sandpile is possible unless particles can leave the system. 
The model can be represented by a directed graph on $N$ vertices, where we
draw $(-\Delta_{ij})$ directed bonds from site $i$ to site $j$, and
$(\sum_j \Delta_{ij})$ arrows from $i$ to outside ( fig. 1). 
\begin{figure} 
\begin{center} 
\leavevmode
\psfig{figure=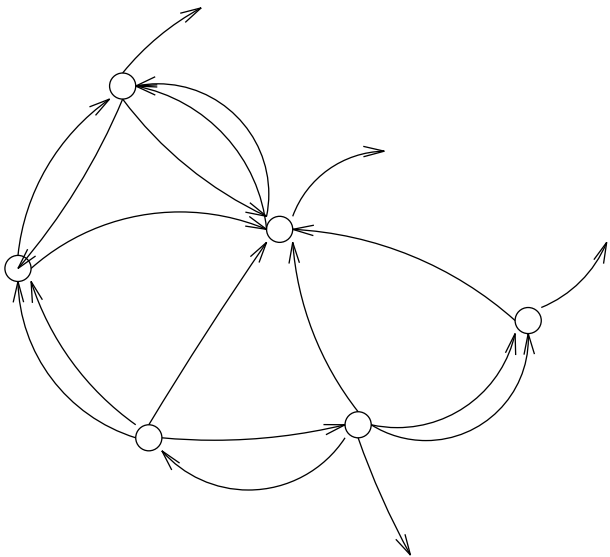,width=8cm,angle=0} 
\caption{ A graphical representation of the general ASM. Each node denotes
a site. On toppling at any site, one particle is transferred along each
arrow directed outward from the site.} 
\end{center} 
\end{figure}

Without loss of generality, we can assume that $z_{ic} = \Delta_{ii}$. 
Then if the site $i$ is stable, we have $0 \leq z_i < z_{ic}$. We
start with a stable configuration of the pile, and add a particle at
random. If this leads to unstable site, it is relaxed using the toppling
rule. If this makes some other sites unstable, they are toppled in turn
until a stable configuration of pile is reached. then we add another
grain, and repeat the process. After a large number of grains are added,
the system loses memory of the initial state, and reaches a statistically
stationary state. In this stationary state, relaxation after the addition
of another grain typically involves a sequence of topplings. This is
called an avalanche. The size of avalanche is a random variable. In many
cases of interest, it seems to have a power law tail, which is a signal of
existence of long-ranged correlations in the system. 

We may use different measures to determine the size of an avalanche. These
are the total number of topplings $s$, the number of distinct sites
toppled $s_d$, the diameter of the region affected by avalanche $R$, the
duration of the avalanche $t$. The probability that the avalanche has
`size' $x$ in the thermodynamic limit of large system sizes will be
defined to vary as $x^{-\tau}$. The exponents $\tau_s, \tau_d, \tau_r$
and $\tau_t$ will be used for the size measures $ s, s_d, R, t$
respectively.   

This model has a very important abelian property \cite{dd90}.  We define
operators $a_i$, which act on the space of stable configurations of the
model. If $C$ is any stable configuration,  $a_i C$ is the stable
configuration obtained by adding a particle at $i$, and relaxing the
system. It is easy to check that given any unstable configuration with two
or more unstable sites, we get the same configuration by toppling at an
unstable site $i$, and then at unstable site $i'$, as we would get if we
toppled first at $i'$, and then at $i$.  Thus the topplings commute with
other. Also, the process of addition of particles commutes with topplings.
By repeated use of these properties, we conclude that the operators
$a_i$'s commute with each other. 

\beq
[a_i, a_j] = 0 , ~~~for ~ all ~i ~and ~j.
\eeq

Adding $\Delta_{ii}$ particles at site $i$ will necessarily cause a
toppling there, whatever the configuration. Thus, it is same as adding
$-\Delta_{ij}$ particles at all sites $j \neq i$. Thus the operators $a_i$
satisfy the equations

\beq
\prod_{j=1}^N a_j^{\Delta_{ij}} = 1,~~ for~ all~ i.
\eeq

Because of the randomness in where grains are added, the time evolution in
this model is Markovian. As in standard Markov theory, the stable
configurations of the sandpile can be divided into two classes: recurrent
and transient. Let us denote by ${\cal R}$ the space of recurrent
configurations.Then ${\cal R}$ is closed under multiplication by
$\{a_i\}$.  In ${\cal R}$, we can define an inverse operator $a_i^{-1}$
for each $i$.  Then restricted to ${\cal R}$, the operators $\{ a_i\}$
form a finite abelian group \cite{dd90}. 

We define an equivalence relation in the space of all configurations (
stable, transient or  unstable) by the property that $\{z_i\}$ and $\{
z_i'\}$
are equivalent under toppling if there exist integers $\{n_i\}$ such that
\beq
z_j = z_j' + \sum_i n_i \Delta_{ij} , for ~~ all ~j.
\eeq
If $\{z_i\}$ are represented as lattice points in an N-dimensional
euclidean space, the equivalent points form a superlattice. 
%%%%%%%%%%%Fig 2 somewhere here
\begin{figure}
\begin{center}
\leavevmode
\psfig{figure=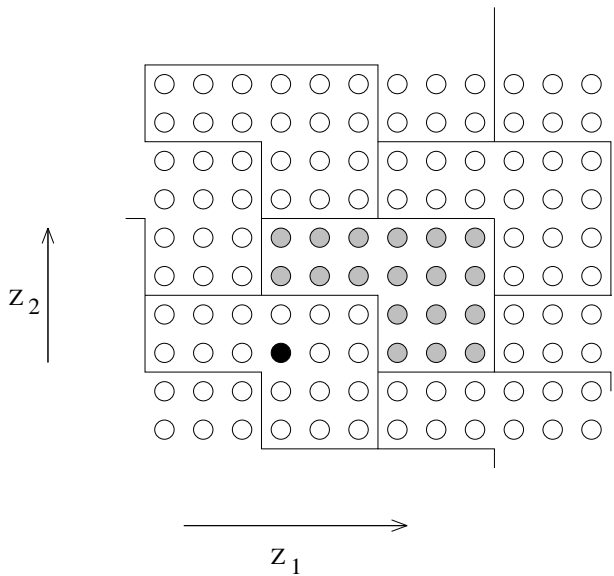,width=8cm,angle=270}
\caption{A tiling of the 2-dimensional lattice of all configurations
$(z_1,z_2)$ for a 2-site automata with $\Delta =
\left[\protect\begin{array}{rr} 6 & -2 \protect\\ -3 & 4
\protect\end{array}\right]$ with copies of the recurrent set $\cal R$. The
grey vertices are the recurrent
configurations.  The black circle marks the configuration $(0,0)$.}
\end{center}
\end{figure}
The
basis-vectors of the superlattice are the rows of the matrix $\Delta$.
It can be shown that there is exactly one recurrent configuration in each
such equivalence class. Hence we get, $|{\cal R}|$ equals the volume of a
unit cell of  this superlattice. Hence we get
\beq
|{\cal R}| = Det \Delta
\eeq
The shape of the recurrent set $\cal R$ is in general quite nontrivial. A
simple case $N =2 $ is shown in fig. 2.

The state of the system after addition of $t$ grains to the system
(``time" $t$) is specified by a probability vector
\beq
|P(t)> = \sum_{C \in {\cal R}} Prob(C,t)|C>
\eeq
This evolves according to the equation
\beq 
|P(t+1)>~= ~~\lgroup \sum_{i=1}^N p_i a_i \rgroup ~~|P(t)>
\eeq
Since the operators $\{a_i\}$ commute with each other, these can be
simultaneously diagonalized. Let the vector $|\psi>$ be simultaneous
eigenvector with
\beq
a_i|\psi> = exp(i\phi_i) |\psi>,~~for ~~all ~~i.
\eeq
Then $\{\phi_i\}$ using eq.(3) are found to be expressible as
\beq
 \phi_i = 2 \pi\sum_j [\Delta^{-1}]_{ij} n_j, ~~for ~~all ~~i.
\eeq
where $\{n_i\}$'s are some integers. Different choices of $\{n_i\}$ give
different eigenvectors. The choice of all $n_i=0$ 
gives $a_i=1$ for
all $i$. This corresponds to the steady state. It follows that in the
steady state all recurrent configurations occur with equal probability.
This characterizes completely the steady state of the system. Since the
time evolution operator is a linear combination of different $a_i$'s, that
is also completely diagonalized. From Eq.(9)  one
can show that for a d-dimensional lattice of size L with nearest neighbor 
topplings, the longest relaxation time
scales as $L^d$. 

There is a simple algorithm (called the burning algorithm) to test if a
configuration is recurrent or not. We scan the lattice, and recursively
`burn' any site $i$ for which $z_i \geq \sum_j' (-\Delta_{ji})$,
where the primed sum is over all unburnt sites $j$. If
eventually, all sites are not burnt away, the configuration is transient.
The unburnt sites form a forbidden subconfiguration (FSC). The
simplest example of an FSC for an ASM on a
d-dimensional hypercubical lattice is two adjacent sites
with both having height 0. If the matrix $\Delta$ is symmetric, and all
sites burn away, the configuration is always recurrent. For non-symmetric
$\Delta$, in which some sites have higher indegree than outdegree,
there may be transient configuration that pass the burning test. Then one
needs a more stringent test, called the `script' test \cite{speer}.

We can also determine some correlation functions in the steady state. If
$G_{ij}$ is the average number of topplings at site $j$ due to a single
particle added at $i$ in the steady state, the condition of mass balance
gives \cite{dd90}
\beq
G_{ij} =  [ \Delta ^{-1}]_{ij}
\eeq

The algebra of operators $\{a_i\}$ has an interesting relationship to the
familiar quantum mechanics. Here
the set of basis vectors of the Hilbert space is the space ${\cal R}$
of recurrent configurations, or equivalently, the space of all
configurations $\{z_i\}$, with configurations equivalent under toppling
identified. The operators $a_i$ are translation operators in this space, (
like $exp(ip)$ ) and they commute with each other. One can generalize
these operators to define operators $a_i(\epsilon)$, which add a phase
factor $\epsilon$ each time a toppling occurs. This generalization
preserves the abelian property. In fact, the evolution of $a_i(\epsilon)$
with the parameter $\epsilon$ is governed by a `quantum-Hamiltonian' which
is linear in the position operators $\{z_i\}$ \cite {extended}. The
nontrivial nature of avalanches is reflected in the fact that commutation
relations between $\{z_i\}$ and $\{a_i\}$ are complicated, and difficult
to write down. 

The mathematical structure of the abelian group for a general finite $N
\times N$ matrix $\Delta$ has also been investigated. In general, a finite
abelian group ${\cal G}$ can be expressed as a product of cyclic groups
$~Z_{d_1}\times~Z_{d_2}\times \ldots Z_{d_g}$, where $g$ is the minimum
number of generators of ${\cal G}$, and $d_i$ is a multiple of $d_{i+1}$.
To determine these integers $\{d_i\}$, we express the integer matrix
$\Delta$ in its normal form (this can be always done) 
\beq
\Delta = A D B
\eeq
where $A$ and $B$ are integer matrices of determinant $\pm 1$, and $D$ is
diagonal integer matrix with diagonal entries $d_1,~d_2,~d_3...$. We can
also choose $A$ and $B$ so that $d_i$ is a multiple of $d_{i+1}$.  The
eigenvalues of $D$ are precisely the
cycle-lengths $d_1,~d_2,\ldots ~d_g$. The remaining $N-d_g$ eigenvalues of
$D$ are $1$. The generators of these sub-groups of ${\cal G}$ can be
explicitly written down as a product of powers of $\{a_i\}$'s in terms of
the (non-unique)  matrices $A$ and $B$. The number of generators is a
complicated function of $\Delta$. In the special case of two-dimensional
BTW model, it can be shown that the the number of generators for a $L
\times L $ lattice is $L$ \cite{algebraic}.

\section{ Directed Abelian Sandpiles}

The simplest of the sandpile models is when particle transfer occurs
preferentially in one direction ( say provided by gravity). If there
is no transfer in the reverse direction, the matrix $\Delta$ becomes upper
triangular. Then $Det \Delta = ~\prod_{i=1}^N \Delta_{ii}$.  Thus all
stable configurations are
recurrent (there are no forbidden subconfigurations).
Correspondingly, the
measure of different configurations in the steady state is the simple
product measure, and averages of various quantities in the steady state
are
easy to compute \cite{d2r2,kathmandu}.

Another simplification which occurs due to directedness is
the fact that for most cases of interest( say, d-dimensional
hypercubical lattice) each site can topple at most once in an avalanche.
Thus we have no multiple topplings, and $s= s_d$. Also, the duration of
the avalanche is equal to its longitudinal size. These simple facts reduce
the number of unknown independent exponents in the model to 2.
%%%%%%%%%%%%%%%%Fig3 somewhere here
\begin{figure}
\begin{center}
\leavevmode
\psfig{figure=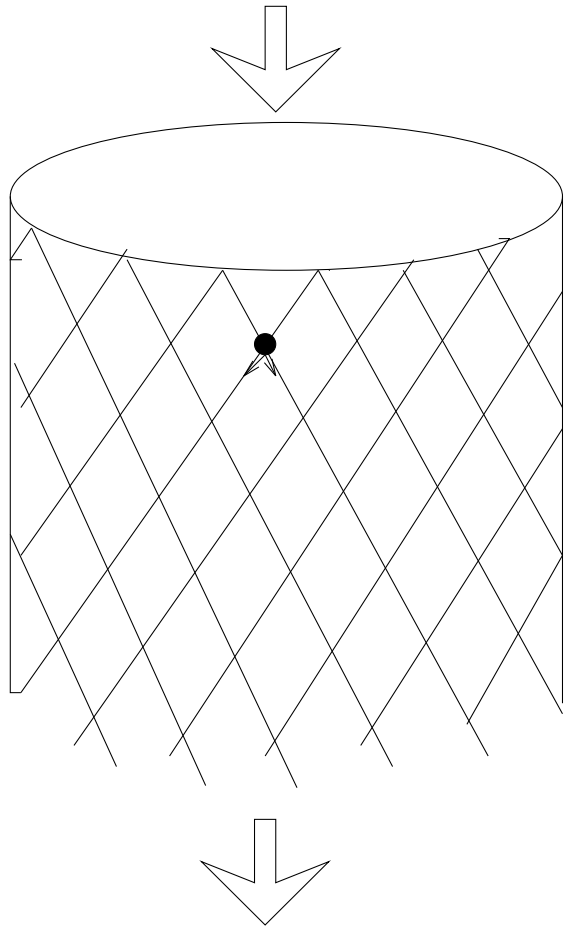,width=8cm,angle=0}
\caption{ A two dimensional directed ASM. Particles are added on top, and
removed from the bottom. On toppling at any site, two particles are
transferred downward.}
\end{center}
\end{figure}

It is straight forward to determine the distribution of avalanche sizes 
in $d$ dimensions. We consider the
ASM on a d-dimensional hypercubical lattice. We define the longitudinal
coordinate of site ${\bf X} \equiv (x_1,x_2,..x_d)$ as $T=\sum x_i$. We
consider a finite lattice with $ 0 \leq T < L$. We assume periodic
boundary conditions in the transverse direction. The pile will be
assumed
to be driven by particles added at top layer ($T=0$), and removed from the
bottom layer ($T=L-1$). The matrix $\Delta$ is assumed to have
$\Delta_{ii}$=d for all sites $i$, $\Delta_{ij}=-1$, if $j$ is a nearest
neighbor of $i$ in the direction of increasing $T$ (fig.3). 

Let $G_0({\bf X;Y})$ be the probability that if a particle is added at the
site ${\bf Y}$ , it will cause a toppling at the site ${\bf X}$. Since in
the steady
state, the height $z_{\bf X}$ takes values $0,1 \ldots (d-1)$. Thus, the
probability that a site topples, given that $r$ of its backward neighbors
have toppled is $r/d$. This implies that $G_0({\bf X;Y})$
satisfies a linear equation
\beq
G_0({\bf X;Y}) = \frac{1}{d} \lgroup \sum_{i=1}^d G_0({\bf X -e}_i;{\bf
Y})
+ \delta_{\bf X,Y} \rgroup
\eeq
Such stochastic processes are well studied in literature under the name of
voter models \cite{voter}. The explicit solution of the above equation is
easily written down:
\beq
 G_0({\bf X;0}) =  d^{-1-T} \lgroup T({\bf X})! /\prod_i X_i! \rgroup
\eeq
For large $X$, this has the familiar diffusive form
\beq
G_0 \sim  T^{\frac {-d+1}{2} } exp ( - Const. R_{\perp}^2/ T),
\eeq
where $R_{\perp}$ is the length of the transverse component of $X$.

Let $m(t)$ be the expected number of sites which topple on the surface
$T=t$ given that there is at least one toppling on this layer.  The
probability that the avalanche reaches upto layer $T=t$, by definition, 
varies as
$t^{1-\tau_t}$. As the average flux out of this surface in the steady
state must be 1 particle per avalanche, we get $m(t) \sim t^{\tau_t-1}$.
As the average mass of an avalanche cluster of duration $T$ is 
$\int_0^T m(t)dt$, we see that the number of sites toppled in the
avalanche $s$ scales as $ t^{\tau_t}$. This immediately gives
\beq
\tau_s = \tau_d = 2 - 1/\tau_t
\eeq

We  define the 3-point function $G({\bf X,Y;0})$ as the probability that
toppling occurs at both sites ${\bf X}$ and ${\bf Y}$, when a particle is
added at ${\bf 0}$. We shall restrict ourselves to the case when $T({\bf
X})= T({\bf Y})$. In this case, this function also satisfies a difference
equation, and the equations for 3-point function does not involve higher
order n-point functions. It is easily shown
that it satisfies also a linear equation
\beq
G({\bf X,\bf Y;0}) = \frac{1}{d^2} \sum_i \sum_j G({\bf X-e_i,\bf
Y-e_j;0});
~~for ~~{\bf X} \neq {\bf Y}.
\eeq
where ${\bf e_i}$ and ${\bf e_j}$ are the
unit vectors of the d-dimensional lattice. For ${\bf X=Y}$, we have the
obvious condition that acts as boundary condition for the previous
difference equation:
\beq
G({\bf X,X;0}) = G_0({\bf X;0})
\eeq
These equations can be solved  recursively \cite{d2r2}.
Summing $G({\bf X,Y;0})$ over ${\bf X,Y}$, we get $<F^2>$, where $F$ is 
the number
of site toppled at the layer $T=t$. It is found that this varies as
$t^{1/2}$ for $d=2$. For $d=3$, one finds $<F^2> \sim t/ (log t)$.
This logarithmic
correction factor to the power-law behavior has been checked in recent
numerical simulations \cite{lubeck}. For $d>3$, the
variance varies as $t$. Thus we see that
\beq
\tau_t = 3/2, for ~~d=2
\eeq
\beq
 ~~~~~~~ = 2 , for ~~d \geq 3.
\eeq
And the transverse size of the cluster at layer $t$ varies as $t^{1/2}$
in all dimensions. Using scaling relations, all other
avalanche
exponents can be determined. For the special case of two dimensions,  the
avalanche clusters have no holes, and the result
$\tau_t =3/2$ can be obtained by using the fact that the left and right 
boundaries of the cluster can be thought of as annihilating random
walkers on a line. 

However, the compactness of the avalanche clusters is
not a necessary condition for the validity of this result. The
discussion above is easily adapted to other cases. For example, consider
the 2-dimensional model, in which the toppling rule is that on toppling
at site $(x,y)$, we transfer one particle each to the sites $(x+1,y-1),
(x,y-1)$, and $(x+1,y-1)$. For this model also, the avalanche exponents
are same as for the simpler cases discussed in \cite{d2r2}. 

The d-dimensional directed ASM turns out to be equivalent to an
(d-1)-dimensional aggregation model with
uniform injection of particles proposed by Takayasu \cite{takayasu}.
The $d=2$ case was proposed by Scheideggar as a model of river-networks
\cite{rivers}. In
the Takayasu model, at each time step, one particle of
mass $1$ is injected at
each site. Then each particle jumps to a neighbor chosen at random. If
more than one particles jump to a site, they coalesce, and their masses
add. After a long time, the distribution of masses tends to  a limiting
distribution. In a space-time picture, if we draw arrows along the
direction of particle-diffusion, we get a directed spanning tree (which
can be thought of a picture of a river network). There is a
one-one correspondence between different trees, and configurations of the
ASM. It is easily shown that on removing a randomly chosen site, the
probability that removed subtree is $S$, is the same as the probability
that the avalanche cluster of ASM is $S$, for all shapes and sizes of $S$.
In particular,  in the steady state of the Takayasu model, the probability
that the mass at a randomly chosen site 
is $m$ is exactly the same as the probability that there will be $m$
topplings in the
corresponding ASM in one avalanche.

 In fact, we can easily adapt the analysis to even fractal
lattices, so
long as there is a translational invariance in the longitudinal direction.
The prototype of such lattices is the so called toblerone lattice,
obtained by a direct product of a linear chain with the Sierpinski gasket
\cite{toblerone}. The details of calculation are omitted here. One finds
that $<F^2> \sim S(t)$, where $S(t)$ is the expected number of distinct
sites visited by  a random walker up to time $t$ on the base fractal (
here the Sierpinski gasket). As $S(t)$ is known to
vary as $t^{d_s/2}$, where $d_s$ is the spectral dimension of the base 
fractal, we get that in general 
\beq
\tau_t = d_s/2 +1, for~~ d_s~ \leq ~2.
\eeq

\section{Undirected Sandpiles}

Since for the directed ASM's, all the critical exponents can be
determined exactly,
these are more suited for pedagogical purposes than the
original undirected BTW model. However, the undirected BTW model continues
to tantalize researchers. It
been studied a lot by theory, and simulations, and we shall now
summarize our current understanding of it.

\subsection {One-dimensional sandpiles}
The simplest undirected lattice is the 1-dimensional linear chain. This
simple case was already solved by BTW in their first paper \cite{BTW1}.
In this case, it is easy to see that on a linear chain of size $L$, with
$z_{ic}=2$, and toppling to nearest neighbors, we get $det \Delta = L+1$.
Only configurations with at most one site having height $0$ are recurrent. 
Almost all avalanches are large, and the $Prob(s < L^{\alpha})$ goes to
zero as $L$ is increased for all $\alpha < 2$. The probability of
avalanche having $s$ topplings has the scaling form
\beq
Prob(s;L) \simeq  L^{-2} f(s/L^2)
\eeq
The detailed form of the scaling function $f$ has been worked out by
Ruelle and Sen \cite{ruelle}. 

A natural question is if this behavior is generic to one-dimensional
models. We have studied this question by analysing the behavior of ASM's
on more general one-dimensional graphs, with more complicated unit cells
\cite{ali1,ali2}.  Interestingly, it was found that the linear chain is
atypical, and in general, on  1-dimesional lattices with more complicated 
unit cells, one finds two types
of avalanches (fig. 4). In the first type of avalanches, $s$ scales
linearly with
$R$. In the second type, $s \sim R^2$. Both these types occur with
comparable frequencies.  One cannot find a simple finite-size scaling form
to describe the distribution of avalanche sizes, but needs a more
complicated linear combination of two simple scaling forms (LC2SSF) 

%%%%%%%%%%%%%Fig 4 somewhere here
\begin{figure}
\begin{center}
\leavevmode
\psfig{figure=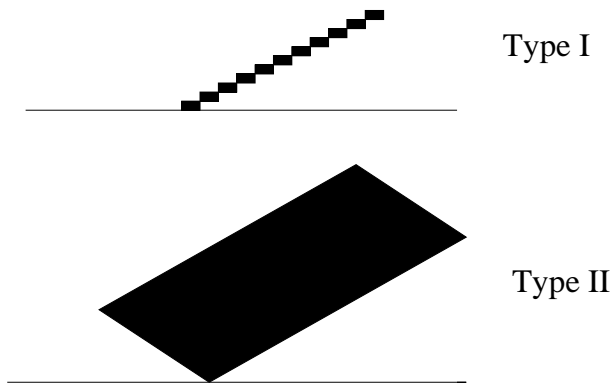,width=8cm,angle=270}
\caption{ Space-time histories of the two types of avalanches for
one-dimensional lattices. The x-direction is space, and y-direction is
time. A box denotes a toppling event.}
\end{center}
\end{figure}

\beq
Prob_L(s) \sim ~~ L^{-1} F_1( s/L) ~+~  L^{-2} F_2(s/L^2)
\eeq
Kadanoff et al had noted that simple finite-size scaling does not 
seem to work for a large class of 1-d models of SOC \cite{KNWZ}. They
proposed a multifractal description of avalanches, which corresponds to an
integral over finite-size scaling forms with different scaling exponents.
The LC2SSF form is simpler as it has only a finite sum, and not an
integral. 
Recently, De Menech et al have argued that even in two dimensions, the
simple finite-size scaling does not work for the distribution of avalanche
sizes in the ASM \cite{demenech}. This is an indication that the two
dimensional avalanches may also be described by an LC2SSF form.

\subsection{ Sandpiles in very large dimensions}

In large dimensions, one expects that the mean-field description of
sandpiles will be correct. Mean-field descriptions of sandpiles have been
tried in several different ways by different authors. The earliest
treatment is by Tang and Bak \cite{tang}. For other approaches, see
\cite{ivashrev, gaveau, janowski, flybjerg}. The simplest treatment is to
treat the
sandpile model on a Bethe lattice. The undirected ASM on a Bethe lattice
was solved exactly in \cite{bethe}.  

The technique involves writing exact recursion equations for the
probabilities of sub-configurations on a given part of a tree. For
example, one finds that for a 3-coordinated tree, far away from the
boundary, the heights $z= 0, 1, 2$ occur with probabilities $\frac{1}{12},
\frac{4}{12}, \frac{7}{12}$ respectively. One can also determine
correlations between
heights at different sites. It is found that the probability of an
avalanche of size $s \sim ~~s^{-3/2}$, for large $s$.  The probability
that the duration of avalanche is $T$ varies as $T^{-2}$ for large $T$.
Multiple topplings are rare, and the probability of $n$ topplings at
the origin in the same avalanche decreases as $exp(-exp(n))$.

These exponents are the same as for critical percolation clusters on the
Bethe lattice, and confirm our expectation that the avalanche process on
this lattice propagates as a critical infection process. The
mathematical treatment and the connection to
the critical infection process is much easier to see if one considers a
directed Bethe lattice , with equal number of arrows in and out at each
site \cite{peng}. 

\subsection{Equivalence to Spanning Trees and to $q\rightarrow 0$ Potts'
Model}

An ASM with a given symmetric toppling matrix $\Delta$ can be represented
by an undirected  graph
with $N+1$ vertices, corresponding to the $N$ vertices of the ASM, and one
additional vertex called the sink. We construct this graph by putting
$(-\Delta_{ij})$ edges between
the vertices $i$ and $j$, and $\sum_j \Delta_{ij}$ edges between site $i$
and the sink. From the well-known matrix-tree theorem \cite{tree}, it
follows that the number of spanning trees on this graph = $Det ~\Delta$,
which is also $|{\cal R}|$.

A one-to-one correspondence between the recurrent configurations of ASM
and the spanning trees can be set up by using the order in which `fire'
propagates in the burning algorithm in  a configuration $C$ to construct
the spanning tree corresponding to $C$ \cite{0-potts}. 

In the burning algorithm, a site with height $0$ is burnt only after all
its neighbors are burnt. In the corresponding spanning tree, such  a site
is  a leaf-site. The fraction of sites which have $z_i=0$ thus gets
related
to the fraction of sites that are leaf sites in a large spanning tree
\cite{ snmddheights}.
For a square lattice, the latter fraction is calculable, and gives
the concenttration of sites with $z=0$ on the square lattice BTW model 
as $f_0 = \frac{2}{\pi^2} \lgroup 1 - \frac{2}{\pi} \rgroup$. One can
easily calculate the fraction of sites of higher coordination numbers
in the spanning tree problem, but that is not directly related to
probabilities of heights in the ASM case. The calculation of the latter  
is much more
difficult, and requires somewhat sophisticated combinatorial graph theory,
and the calculation has only been done for the square lattice so far
\cite{heights}. 

One can also calculate the probability that two sites at distance $R$ from
each other both have heights $0$. This can be done by a straightforward
generalization of the 1-site problem \cite{snmddheights}.  It is found
that in d-dimensions, this probability for large $R$ varies as $ A + B
R^{-2d}$.  It is expected that similar $R^{-2d}$ tails will be seen in the
joint probabilities for other heights, but the calculation is more
difficult. The $R^{-4}$ behavior of correlations is also seen for the
surface sites in the two-dimensional case \cite{surface}.

The spanning tree problem is well-known to be equivalent to the resistor
network problem, and Fortuin and Kasteleyn showed that it can be
considered as the $q \rightarrow 0$ limit of the $q$-state Potts' model
\cite{wu}. Thus, the undirected ASM is equivalent to the $q=0$ Potts
model, and not to the $q=1$ Potts model, which might have been expected on
the basis of its connection to the percolation problem.
 
\subsection{Exponents of the Two dimensional Model}
 
The equivalence to the equilibrium 2-dimensional Potts model allows one to
use known exponents of the latter to predict exponents for the ASM. 
At its critical point, the q-state Potts model shows additional symmetry:
the conformal invariance. The central charge corresponding to $q=0$ is
$c=~-2$. For this value of central charge, one can look up the Kac table
for other exponents of the theory.  We find that the fractal dimension of
chemical paths in the spanning trees problem in two dimensions is $5/4$
\cite{saleur,coniglio}.
Since the outward propagation of activity in the avalanches, is like the
spread of fire in the burning algorithm, we identify this with the
exponent relating the distance to time of avalanches $T \sim ~R^{5/4}$.
In terms of exponents, this gives the relation
\beq
(\tau_t -1) = \frac{4}{5} (\tau_r -1)
\eeq

Another result which can be derived exactly is the distribution of
avalanche sizes that are formed by adding a particle at the vertex of a 
wedge of angle $\theta$ with open boundaries\cite{wedge}. In this case,
there is only one wave of toppling. 
Hence probability of avalanche reaching distance $R$ is same as the
propagator $\Delta^{-1}$. This  is easily evaluated by using complex
cordinates. One finds that the probability decreases as $R^{-x}$, with
$x~= ~ \pi/\theta$. Using the compactness of clusters, one finds that
for a wedge of angle $\theta$ the value of corner exponent $\tau_{corner}$
is $1+ \frac {\pi}{2 \theta}$. For $\theta = 2 \pi$, we get
$\tau_{corner}=5/4$.
One may expect that adding a particle at the end of an open half-line
should
give rise to fewer
topplings than adding a particle in bulk. This would suggest that
$\tau_d \leq 5/4$ in two dimensions. Unfortunately, it is difficult
to
prove such an inequality because of the many more recurrent states in the 
wedge problem than in the bulk case, and it is difficult to compare the
probabilities of different events. 

\subsection{ Waves of toppling}

An avalanche in the ASM can be broken into a sequence of sub-avalanches.
Let us call the site where a new particle is added  $O$.
Since the topplings can be done in any order, we topple once at $O$, and
then topple any other unstable sites. This disturbance spreads as a wave,
called the first wave of toppling. If $O$ is still unstable, we topple
once again at $O$ and let other sites relax, constituting the second wave
of toppling. This process is continued until $O$ becomes stable. Thus, an
avalanche is broken into a sequence of waves of toppling \cite{waves}.
It was shown by Priezzhev that the set of all waves is in one-to-one
correspondence with all two-rooted spanning trees. From the fact that in
two dimensions, the propagator $\Delta^{-1}$ has a logarithmic dependence
on distance $R$, he showed that in an ensemble in which all waves have
equal weight ( an avalanche with $n$ waves is counted $n$ times), one gets
\beq
Prob(wave~ with~ s ~topplings) \sim ~~1/s, ~~for  ~s >> 1.
\eeq
Using the fact that the distribution of sizes of last wave is related to
the distribution of number of sites disconnected from a tree if a
randomly chosen site is removed from a spanning tree, and the known
fractal dimension of the paths along spanning trees, it can be shown that
the probability of the size of the last wave is $s$ varies as $s^{-11/8}$
for large $s$ \cite{inverseaval}.

\subsection{ Recent Developments}

Priezzhev et al \cite{pki96}  developed a scaling picture of
avalanches in the two-dimensional undirected ASM, based on the
decomposition of avalaches into waves of topplings. The basic
picture of Priezzhev et al depends on observation that the progress of an
avalanche typically depends on a fast expansion, and then slow
contraction. In the contraction phase, if the k-th wave is of size $s_k$,
the next wave is smaller, with $s_k - s_{k+1}$ being of order $s_k ^x$,
with $x < 1$.  A cluster of size largest wave-size $ s_d$, would
typically have $n_c \sim s_d^{1-x}$ waves, and hence $s \sim s_d^{2-x}$.
Using some plausibility arguments for the value $ x=3/4$,
these authors conjecture that the exact values of the exponents are 
$\tau_d = 5/4, \tau_s =6/5$, and $\tau_t= 7/5$.

Paczuski and Boettcher \cite{pacz} have questioned some of the
assumptions used by Priezzhev et al, in particular, the fact that quite
often $s_{k+1} > s_k$.  [Some of these objections have been taken into
account in a revised version of their arguments by Ktitarev and Priezzhev
\cite {ktit}.] Paczuski and Boettcher studied the conditional
probability $Prob( s_{k+1}|s_k)$ that the $(k+1)$-th wave is of size
$s_{k+1}$, given that the previous wave was of size $s_k$ by simulations.
They presented numerical evidence that this quantity varies as
$s_{k+1}^{-a}$ for $s_{k+1} <<s_k$, and as a different power for
$s_{k+1}>>s_k$. They also proposed that this conditional probability is
only a function of the ratio of $s_{k+1}$ and $s_k$, and has the scaling
form

\beq
Prob(s_{k+1}=s'~| ~s_k =s) \simeq (1/s') F(s'/s)
\eeq

However, note that if the scaling variable is $s'/s$, then the change in
$s$ is of order $s$, hence necessarily $x=1$. This makes $\tau_s = \tau_d
=1$.

In a recent paper De Menech et al \cite{demenech} have argued that
avalanches that reach the boundary scale differently than those that
don't. This leads to a breakdown of the simple finite-size scaling theory
usually assumed in deriving scaling relations, and in data analysis.  From
analysis of their simulation data, these authors conclude that the events
which reach the boundary occur roughly with probability $L^{-1/2}$ on a
lattice of size $L$. This suggests that the probability $Prob(R \geq x)$
decreasing as $x^{-1/2}$ for $R < L$, and hence gives $\tau_r = 3/2$, and
$\tau_d=5/4$, same as the value conjectured by Priezzhev et al
\cite{pki96}. However, in this range, there are few waves per avalanche,
and De Menech et al argue that $s \simeq s_d$ for $s_d < L^2$. 

The avalanches which reach the boundary are only a small fraction
$(L^{-1/2})$. De Menech et al found that on the average each of these
takes about $L^{1/2}$ more
waves before it is stopped. Thus, for these avalanches, each contraction
is of order $L^{3/2}$, which is also consistent with the Priezzhev et al
proposal that $x = 3/4$. For these avalanches, $ L^2 < s < L^{5/2}$.  If
we argue that the probability that $s$ is of order $L^{5/2}$ is of order
$L^{-1/2}$, this gives $\tau_s = 6/5$, as argued by Priezzhev et al.
However, as the behavior for the full range of $s$ cannot be described by
a single power law, these exponents are only effective exponents.

\bigskip
  
\section{ The  Abelian Distributed Processors Model}

   We would like to distinguish between the properties of the ASM that are
specific to the details of the model, and those that hold for a larger
class of SOC models.  This has motivated us to study variations of the
ASM, which keep some of its abelian structure intact. This is described
below. For a different generalization, see \cite{chaucheng,chau}. In
particular, we show that the some sandplie
models
with stochastic toppling rules ( e.g. the Manna model \cite{manna}) still
satisfies the abelian property \cite{biham}.  The stochasticity in
toppling rules models phenomenologically the variation of grain sizes,
roughness etc. that are found in real experiments on granular
media\cite{ricepile}. In the next section, we shall describe a different
variation which makes the model non-abelian.

  Consider a network of $N$ finite-state automata  $P_i,~~ i=1$ to $N$.  
The number of internal states may be different for different
processors. Each processor is provided with an input stack, which can
store
messages until they are read, and some output channels to communicate to
some of the other computers.
At quasi- regular intervals, each computer checks its own input stack. If
the stack is empty, it waits with no change of state. If the input stack
is not empty, it pops a message from the stack ( acoording to some fixed
local protocol, say first-in-first-out), and reads it.  At each processor
$P_i$,
there is a rule table ${\cal R}_i$ which tells what will be the final
state $\sigma_{fin}$ of the processor $P_i$, and what output messages
and to which processor(s) are generated, given that initial state of
the
processor $\sigma_{ini}$, and the message read \cite{ CRP}.

An input may generate none, one or more  output messages
depending on the initial state of the processor. We may assume, for
simplicity, that each message consists of a single letter from a finite
alphabet. The rule table, as well as, the processor speed can be different
at different nodes of the network. Again, for simplicity, we ignore
transmission delays, and an output generated will be assumed to reach the
input stack of the receiver instantly. Some messages may also be sent
`outside', i.e. to computers outside the system ( say, a human). 
 
In a quiescent state of the network, all the input stacks are empty, and
the configuration of the full network is specified by $\{\sigma_i\}$,
where $\sigma_i$ is the internal state of each processor $P_i$. To begin a
computation, we select an input message $m_{in}$ at random, drawn from
some given fixed distribution and put the message in the input stack of
one of the processors $P_{i}$, also chosen at random. This will cause the
processor to change its internal state, and may generate messages to other
processors. These new processors are now activated, and the computation
goes on. Eventually, the system would again get into a state where all
input stacks are empty, and no further processing occurs.  This will be
taken as the end of the computation.

We restrict our attention to systems where all computations halt after a
finite number of steps.  Once, an outside observer finds that the
computation has stopped \cite{observer}, he/she can reactivate the system
by choosing again a site at random, and sending a message to its input
stack.

 The evolution of this network from one quiescent state to another is
clearly Markovian, and after a large number of messages have been
processed, it reaches a stochastic stationary state. This state is
characterized by its probability distribution $Prob(\{\sigma_i\})$, of
different quiescent states $\{\sigma_i\}$. Sending a message to a
processor
initiates a {\it computation}. The {\it size} of computation may be
measured in terms of number of messages generated, or of processors
affected, etc.. In the context of self-organized criticality, we would be
interested in systems where such distributions show power-law tails. 

      In general, the result of such a computation will depend on the
order in which different messages are received at a particular processor. 
In information science, one is often interested in ensuring that
the network computation is independent of the speeds of
different processers. This often requires setting up of controls by
which some
messages have to wait till some others have been read etc.. In the
following, we restrict ourselves to computations in which the final result
is always independent of the order in which inputs are received. 
      
We assume that the rule-tables at each of the processors $P_i$  are such
that for all
initial states $\sigma_i$, and all input messages $m$ and $m'$, the final
state $\sigma_f$ and the set $\{o_j\}$ of output messages generated is the
same as would result if the messages were read in the reverse order (
first $m'$, and then $m$). Thus we assume the abelian property at the
level of individual processors. The validity of
the abelian property can be checked by examining the rule table for
each processor. Clearly, a network in which each component processor is
abelian, is also abelian. We shall
call such networks Abelian Distributed Processors (ADP).

As in the ASM case, we can define mutually commuting operators $a_{i,m}$ 
corresponding to adding the message $m$ to
the processor $P_i$ in a quiescent state $C$ of the network, so that the
resulting quiescent state is $a_{i,m} C$. Restricted to the set of
recurrent states of the network, these operators define an Abelian group.
Clearly, the ASM is a special case of ADP model, where the state
$\sigma_i$ is specified by the height variable, a message is a sandgrain,
and rule table are the toppling rules of ASM.
Many of 
the
results obtained for the ASM are immediately generalized to the ADP case.
It is useful to illustrate the generality of this structure with the help
of some examples.

\subsection{Block-renormalized Sandpiles}
%%%%%%%%%%%Fig5 somewhere here
\begin{figure}
\begin{center}
\leavevmode
\psfig{figure=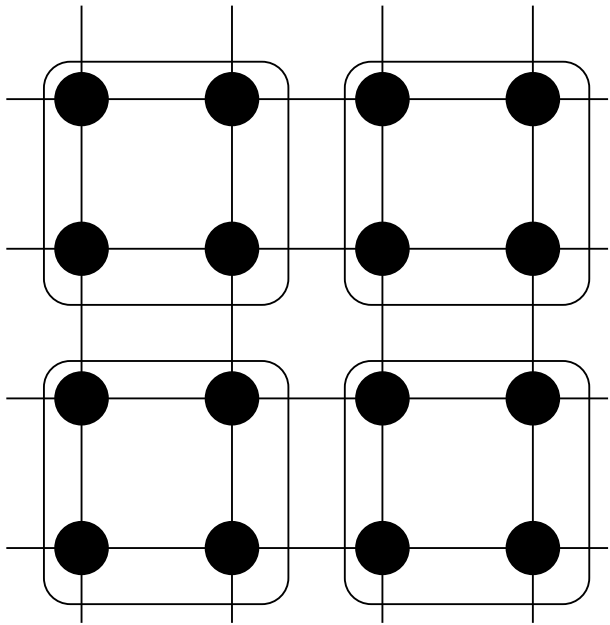,width=8cm,angle=270}
\caption{Block -renormalization of the ASM on a square lattice.}
\end{center}
\end{figure}

  Consider the ASM, say on a square lattice with the usual toppling rule
to nearest-neighbors. We try real space renormalization on this system,
and group the sites together into blocks of size $2 \times 2$ and think of
each block as a single processor (fig. 5).  The internal state of this
block is characterized by four integers $(z_1,z_2,z_3,z_4)$. Adding a
grain could be to any of the four sites in the box. Accordingly, we define
that an input message to the box is a single letter from a four letter
alphabet. Because of forbidden configuration, the different possible
internal states of the box are only $192$, and not $4^4=256$. The rule
table for transitions for each box is easily written down from the
toppling rules of the original model. It is easy to see that this block
renormalized ASM still satisfies the abelian property, and is an ADP
model. 

There has been some recent interest in applying real-space renormalization
group techniques to ASM's ( on regular lattices \cite{RSRG1}, and
also on fractals \cite{fractals1,fractals2}. While these seem to give
rather good estimate of exponents, but the techique involves ad hoc
uncontrolled approximations.  The fact that abelian property is preserved
under renormalization in the ADP model may help in devising more reliable
approximation scemes \cite{RSRG2}.

\subsection{ Eulerian Walkers and Related Models}

We may also consider sandpile models with periodic toppling rules. 
Both the threshold and the transfer rules may vary from one toppling to
next in a deterministic manner {\it independently} at each processor. For
example, consider a sandpile model on a square lattice with toppling
rules as follows: stable heights of the pile are 0 and 1. If height is
greater than 1, two particles are transferred to nearest neighbors. The
transfer is to two neighbors in the vertical direction if in the previous
toppling at that site was to the neighbors in the horizontal direction.
And vice versa.

 The simplest model of this type is the Eulerian walkers model
\cite{Eulerian}. In this model, we imagine a walker on a lattice. at each
site, there is an arrow marking the direction of the last exit of a walker
from that site. A walker is randomly dropped on to a site. At each site,
he resets the arrow to a new direction according to fixed rules,( say by
rotating it clockwise by $90^{\circ}$ if the lattice is the square
lattice),
and then takes a step in the direction of the arrow. At the
boundary, some steps lead to the
walker leaving the system. At this stage, a new walker is introduced in
the system.

 In the steady state, the directions of arrows in the system develop
long-range correlations. In fact, it can be shown that in the quiescent
state after a long time, all the arrow directions will form a
directed spanning
tree. This implies that one can set up a one -to-one correspondence
between the recurrent configurations of the Eulerian walkers model, and of
the corresponding ASM. On closed graphs with no sink sites, the walker
eventually settles into cyclic state. It can be shown that this limit
cycle which visits each bond exactly once. Such a path is called the
Eulerian path, and hence the name of the model. Several properties of this
model can be determined exactly \cite{Eulerian}.  For example, it is found
that on a d-dimensional lattice with an initially random configuration of
arrows, the mean square deviation of the walker from his initial position
increases with time $t$ as $t^{\frac {2}{d+1}}$ for $d \leq 2$, and as $t$
for $d > 2$. However, the distribution of the number of steps taken by the
walker before he leaves the lattice is not known.

 An interesting model in this class of  models with periodic toppling
rules is
obtained if we consider lots of Eulerian walkers
on a lattice, with the rule that a walker arriving at a site just waits
there unless the number of walkers waiting at that site exceeds $q$. If
it does, q
walkers leave in q successive directions, and the arrow is reset. The
ASM corresponds to $q=4$, and Eulerian walkers to $q=1$ \cite{Eulerian}. 
The case $q=2$ on a Bethe lattice of coordination number $4$ has recently
been studied by Shcherbakov  \cite{shcherbakov}. The techniques used
to study the ASM on the Bethe lattice are not easily adapted to this model
as identification of all the forbidden subconfigurations, and
the propagation of avalanches is more complicated.

 The interesting point about these variations is that the particle
addition operators $a_i's$ in this case satisfy the same algebra as the
square lattice ASM [eqs (3)], but the structure and sizes of avalanches is
quite different. In particular, for the Eulerian walkers case, on a
lattice of size $L$, in most cases, a walker takes $O(L^2)$ steps before
leaving the system. Since this is the most natural analogue of number of
topplings in the ASM, in this case, there are no small avalanches.

\subsection{ Abelian Stochastic Sandpile Models}
 The periodic toppling rules of the previous example are most simply
realized by assuming the existence of a finite number $k$ of distinct
rule-tables $R_1, R_2, \ldots R_k$ at each site at each site, and an
infinite sequence $\{r_i\}$ with $ 1 \leq r_i \leq k$, which tells which
of the $k$ rule-tables to follow for the $i-$th toppling. If the sequence
of integers $\{r_i\}$ is periodic, we get the ADP model with periodic
update rules. We may however take more complicated sequences without
losing the abelian character. In particular, we may take this sequence to
be generated by a pseudo-random number generator of a large period. A
system evolving this way, may equally well be described as undergoing
evolution with stochastic update rules. 
 
 Let us consider a square lattice. The stable heights at
each site are assumed to have a value $0$ or $1$. If the height at a site
exceeds $1$, that site topples according to one of the following rules:

 R1: Two particles transferred to north-south neighbors.

 R2: two particles transferred to east-west neighbors.

Each site has a local pseudo-random number generator, which gives a random
single bits (0 or 1) on each request. Before each toppling, a new random
bit is drawn from this local generator. If the output is 0, rule R1 is
used, else R2.  As the output sequence from each generator is actually
deterministic,  this model belongs to the  ADP class. This model is
equivalent to
a model of sandpiles with stochastic update rules first discussed by Manna
\cite{manna}. He studied cases where the critical height is always $2$,
but to which neighbor the transferred particles go was decided randomly.
In his first model, the two particles move in opposite directions with
equal probability, which is equivalent to the ADP discussed above. . In
the second model, for each of the two particles, one of the four neighbors
of the site is chosen at random, and the particle transferred there.

    We define particle addition operators $a_{i,j}$ corresponding to
addition of a particle at site $(i,j)$ and relaxation as before. 
It is
easy to see that these operators still commute:

\beq [a_{i,j},a_{k,l}] =0,~~~for~ all~ i,j,k,l.  \eeq If the internal
states of different random number generators are not accessible to outside
observer, the topplings are effectively stochastic. Applying $a_i$ to a
basis vector, say $|C>$, of the system does not lead to a unique
configuration, but a {\it linear combination of basis vectors}. But the
operators $a_i$ still commute. The closure relations between the operators
are similarly modified. It is easy to see that for Manna's first
stochastic model, these become \beq a_{i,j}^2 = (1/2)( a_{i+1,j} a_{i-1,j}
+ a_{i,j+1} a_{i,j-1})  \eeq for all $1 \leq i,j \leq L$, and assuming
that $a_{i,j}=1$, if $i$ or $j$ equals $0$ or $L$.  If we consider one
dimensional representations of this commutative algebra, we can think of
$a$'s as complex numbers. However, not much can be said at present about
the nontrivial solutions of this set of coupled quadratic equations. Even
the number of nontrivial solutions is not easy to determine. In general, N
simultaneous quadratic equations have $2^N$ solutions. But several of
these may be trivial (all $a$'s zero). For example, we can easily check
that for a 2X2 square, the number of nontrivial solutions of the $4$
coupled quadratic equations in $4$ variables is only $3$. 

Looking at the original problem, it is easy to see that it has some FSC's:
For example, it is easy to show that a 2X2 square with all sites
unoccupied is an FSC. Clearly, there are other larger FSC's. For the small
2X2 square discussed above, the number of recurrent configurations is
easily seen to be $15$.  Thus the number of nontrivial solutions $\{a_i\}$
is no longer equal to the number of recurrent configurations in this
stochastic generalization of the ASM. 

In this respect, Manna's second model is more tractable. In this case,
each of the two particles is moved randomly to one of the four neighbors.
It follows that in this case the operators $\{a_{i,j}\}$ satisfy the
following equation:
\beq
a_{i,j}^2 = \frac{1}{16}[ a_{i+1,j} + a_{i-1,j} + a_{i,j+1} +
a_{i.j-1}]^2
\eeq
These are again reducible to $L^2$ coupled quadratic equations for the the
simultaneous eigenvalues of these operators. But in this case, we can
reduce them to $L^2$ linear equations

\beq
\eta_{i,j}a_{i,j} = \frac{1}{4}[ a_{i+1,j} + a_{i-1,j} + a_{i,j+1} +
a_{i.j-1}]
\eeq
where $\eta_{i,j} = \pm 1$. There are $2^{L^2}$ different choices of the
$L^2$ different $\eta$'s. For each such choice, we get a set of
simultaneous eigenvalues of $\{a_{i,j}\}$. Thus, we get the full set of 
eigenvalues for these operators. Correspondingly, one can show that the
all empty configuration is recurrent in this case. Hence all
configurations are recurrent.

It may be noted that the equations (29) look like the wave-equation of 
quatum-mechanical particle in a random potential. This provides an
intriguing connection to the well-known Anderson localization problem
\cite{anderson}. Here $a_{i,j}$ act like the wavefunction, and
$\eta_{i,j}$ is the random potential.  Of course, in this case, it is an
inhomogeneous linear
equation ( due to presence of boundary terms), and so involves a weighted
sum over all the eigenvalues of the homogeneous case. The sum over the 
eigenvalues of $\{a_{i,j}\}$'s is the average over different realizations 
of the potential in the Anderson problem. 

\section{ A Non-abelian Stochastic Sandpile Model}

This model is a generalization of the directed ASM by making the toppling
rules stochastic in a different way.
The stochasticity is parameterized by
a real parameter $p$, with $0 \leq p \leq 1$. For simplicity,
let us
define it on
the same lattice as in section 3.2. The height at the site $(i,j)$ of the
square lattice is $h(i,j)$. If on adding a particle, the height $h(i,j)$
exceeds $1$, the site becomes {\it unstable}. With probability $p$, the
height at the unstable site decreases by $2$, and one particle is
transferred to each downward neighbor. with probability $(1-p)$, no
transfer takes place. In either case, the site becomes stable for the next
time step. At any time, only sites where at least one particle was added
at the last step can become unstable \cite{bosa}.

The case $p = 1$ corresponds to the directed ASM. For $p \neq 1$, the
model is no longer abelian, as adding two particles at a site one by
one may have a different effect than if they are added together. Also, if
$p \neq 1$, in the steady state of the system, height at a site can take 
arbitrarily large values with a small probability. Interestingly, this
introduction of stochasticity is found
to be a relevant perturbation, and it changes the avalanche exponents. We
determine these exactly in terms of the critical exponents of directed
percolation.  A recent discussion of different universality classes of
sandpile models may be found in \cite{benhur}.

For any site, we can define two parameters $p_1$ and $p_2$, which give the
probabilities that a toppling occurs at the site if $1$ or $2$ of the
sites above it have toppled. The evolution of avalanches in this model is
as in the Domany-Kinzel (DK) model of directed percolation with the two
probabilities $(p_1,p_2)$ \cite{domkin}. In our case, we have $p_2=p$ for
all sites, and
$p_1 = p \rho$, $\rho$ being the probability that the height at the site is
nonzero. In the DK model, there are two phases: for low values
of $p_1,p_2$, all percolation clusters are finite, but for $p_1,p_2$ large
enough, there is a finite probability of an infinite cluster. In our case, 
if no percolation occurs, there is a build up of particles in the top
layers, and no steady state exists.[ This occurs if $p <p^*$, where $p^*$
is the critical probability of directed site percolation on a square
lattice.] Also, $(p_1,p_2)$ cannot be in the
supercritical phase, because then  each layer would lose more paricles
than it gains from the layer above. It follows that the steady state
values of $(p_1,p_2)$ must be very near the critical line of the DK model. 

Thus the local evolution of avalanches in our model must be same as that
of critical cluster in the DK model. But for the critical clusters in the
DK model, it is known that expected number of sites in the $\ell$-th layer
increases as a power of $\ell$. This cannot be satisfied because of the
particle conservation, mean outflux of particle per layer per avalanche is
$1$. The resolution of these conflicting requirements is that the system
in the steady state has a $p_1$ value which varies with $\ell$, as
\beq
 p_1(\ell) = p_{1c} - A \ell ^{-1/\nu_{\|}}
\eeq

where $\nu_{\|}$ is the exponent characterizing the rate at which the
longitudinal correlation length diverges away from critical point $(
\xi_{\|} \sim (\delta p)^{-\nu_{\|}})$. Thus the calculation of
distribution of avalanche-sizes reduces to finding statistics of surface
avalanches when the concentration profile has a power-law dependence away
from surface. In the present case, these exponents depend on the amplitude
$A$. The requirement of mass balance in the steady state thus fixes the
value of $A$, and hence determines the avalanche distribution in terms of
exponents of the directed percolation problem. 

Thus the model with $p \neq 1$ is  a different universality class than
the abelian case $ p=1$. In detail, one finds
\cite{bosa} that in d-dimensions $(1 <d <5)$ the probability that
avalanche
stops at layer $\ell$ varies as $\ell^{-\tau_t}$, where

\beq
\tau_{t} = 1+ (d-1)\nu_{\perp}/\nu_{\|} -\beta/\nu_{\|}.
\eeq
Here $\beta $ and $\nu_{\perp}$ are the standard exponents of directed
percolation.
and the probability that avalanche has exactly $s$ topplings varies as
$s^{-\tau_s}$, with $\tau_s = 2- 1/\tau_t$.

For $d = 2$, using the known estimates of the DP exponents, we get $\tau_s
\simeq 1.313$ and $\tau_t \simeq 1.473$. These differ by only about $2\%$
from the
values $\tau_s =4/3$, and $\tau_t = 3/2$ for the case $p=1$.  The upper
critical dimension changes from the value $d_u = 3$ for $p=1$ to $d_u =5 $
for $p~ \neq ~1$. For all d $~ \geq ~5$, we recover the mean-field values
$ ~\tau_s = ~3/2,~ \tau_t = ~2$. 

Thus all the exponents characterizing avalanche clusters for $p \neq 1$,
are expressed in terms of exponents of directed percolation, but they are
different from the exponents of directed percolation clusters because of
the particle conservation in the model.  The undirected version of this
stochastic model remains unsolved even in $1+1$ dimensions. 

\section{ Concluding Remarks}

 The sandpile is, of course, a non-equilibrium system driven by the slow
addition of grains. One often finds in literature the statement that
non-equilibrium systems are qualitatively different from equilibrium ones,
because one cannot write a `well-behaved' hamiltonian with short-ranged
interactions that will show the often long-ranged correlations observed in
the non-equilibrium steady states(NESS's). But for the ASM, the NESS is
characterized by a Boltzmann measure corresponding to a rather well-known
equilibrium statistical mechanics model( the spanning trees or the q=0
Potts model). For the directed ASM's, the even simpler hamiltonian H=0 is
adequate.

 I think that the conventional wisdom is not really wrong here, because
the transformation of variables needed to go from the height variables of
the ASM to the spanning tree is quite complicated, and nonlocal. It
preserves the probabilities of different configurations in the steady
state, but the simple local rules of toppling in the ASM become
complicated (and ill-understood) nonlocal evolution rules in the spanning
tree description. A description directly in terms of Potts model spins is
not possible because of the the formal nature of the $q\rightarrow 0$
limit.
  
 Another oft-discussed question is related to the fact that in the ASM,
the time-scale of sand addition is assumed to be much longer than the
time-scale of individual topplings. In fact, the existence of two very
widely seperated time-scales has been argued to be  one of the defining
characteristics of self-organized critical systems \cite{grinstein}. But,
as we have seen, the directed ASM in d-dimensions is equivalent to the
Takayasu aggregation model. In the latter model, also driven, there seems
to be no seperation of time-scales:  aggregation and diffusion both occur
at comparable rates!  In addition, the distribution of avalanche-sizes in
the ASM simply translates to the distribution of masses in the steady
state of the second model. 

If we accept the existence of two time-scales as one of the defining
characteristics of SOC, we find ourselves in the uncomfortable situation
that the prototypical model of self-organized criticality is `equivalent'
to a model that does not show self-organized criticality. In this case, we
cannot even resort to the argument about change of variables being
complicated, as the avalanche distribution of ASM simply corresponds to
the mass distribution in the Takayasu model.  We conclude that  the
existence of two widely seperated timescales should not be considered as a
necessary condition for self-organized criticality. A long-time scale can
be defined by the typical seperation between suitably defined rare `burst'
events in the system. In the Takayasu model, these burst events are
formation of large mass clusters. 

Before concluding, let me make a small list of questions which are not
quite well understood yet.

Can one obtain the results for spanning trees, without using the heavy
machinery of conformal field theory? In particular, the result that
tortuisity exponent of paths on spanning trees in two dimensions is 5/4
should have  a direct 'combinatorial' derivation.

What are the exponents of the undirected ASM in two dimensions? 

Is it true that fractal dimension of avalanche clusters is same as space
dimension for all $d \leq 4$? 

Is there a systematic way to exactly compute the probabilities of
avalanches of sizes $1, 2 \ldots $ in the undirected ASM?

What are the time-dependent correlations in avalanche sizes?

I have benefitted a lot from discussions with various colleagues and
friends. Rather than list them all, let me just thank my collaborators A.
A. Ali, A. Dhar, S. Krishnamurthy, S. N. Majumdar, S. S.  Manna, V. B.
Priezzhev, R. Ramaswamy, P. Ruelle,  S. Sen, B. Tadic, and D. N. Verma.
Working with them has been a rewarding learning experience.  
 

\begin{thebibliography}{999}
\bibitem{observer} Clearly, the observer has to be  able to monitor
the status of all the processors in the system.
\end{thebibliography}
\end{document}